\documentclass[aps,
               superscriptaddress,
               twocolumn,
               floatfix,
               pra,
               nofootinbib,
               a4paper]{revtex4-2}
\usepackage{amsmath}
\usepackage{graphicx}
  \graphicspath{{figs/}}
\usepackage[colorlinks=true,
            linkcolor=blue,
            citecolor=blue,
            urlcolor=blue]{hyperref}
\usepackage{natbib}
\usepackage{overpic}
\usepackage[caption=false]{subfig}
\usepackage{times}
\usepackage[usenames,dvipsnames]{xcolor}
  \definecolor{plotgray}{HTML}{AAAAAA}
\usepackage{tikz}
  \usetikzlibrary{arrows.meta,calc}

\begin{document}

\title{Post-quantum nonlocality in the minimal triangle scenario}

\author{Alejandro Pozas-Kerstjens}
\thanks{physics@alexpozas.com}
\affiliation{Institute for Mathematical Sciences - ICMAT (CSIC-UAM-UC3M-UCM), 28049 Madrid, Spain}
\affiliation{Group of Applied Physics, University of Geneva, 1211 Geneva 4, Switzerland}
\affiliation{Constructor University, Geneva, Switzerland}

\author{Antoine Girardin}
\affiliation{Group of Applied Physics, University of Geneva, 1211 Geneva 4, Switzerland}

\author{Tam\'as Kriv\'achy}
\affiliation{Atominstitut, Technische Universität Wien, 1020 Vienna, Austria}

\author{Armin Tavakoli}
\affiliation{Physics Department, Lund University, Box 118, 22100 Lund, Sweden}

\author{Nicolas Gisin}
\affiliation{Group of Applied Physics, University of Geneva, 1211 Geneva 4, Switzerland}
\affiliation{Constructor University, Geneva, Switzerland}

\begin{abstract}
We investigate network nonlocality in the triangle scenario when all three parties have no input and binary outputs.
Through an explicit example, we prove that this minimal scenario supports nonlocal correlations compatible with no-signaling and independence of the three sources, but not with realisations based on independent quantum or classical sources.
This nonlocality is robust to noise.
Moreover, we identify the equivalent to a Popescu-Rohrlich box in the minimal triangle scenario.
\end{abstract}

\maketitle

\section{Introduction}
The study of nonlocality has historically focused on the scenario of the Einstein-Podolsky-Rosen paradox and Bell's inequality.
In this so-called Bell scenario, a single source emits pairs of particles that are  distributed between two parties who independently measure them in spacelike separated locations and then compare their statistics.
In more recent years, the study of nonlocality has moved beyond the Bell scenario to consider the correlations that can arise in network scenarios \cite{ReviewPaper}.
A network features several independent sources that emit particles that are then distributed between a number of parties according to the specific network architecture.
For example, the simplest network, known as the bilocal scenario \cite{Branciard2010,Branciard2012}, has two independent sources that each distribute a pair of particles; one between parties Alice and Bob and one between parties Bob and Charlie. This is the same scenario encountered in the simplest form of entanglement swapping \cite{Zukowski1993}.

It is well-known that the introduction of multiple independent sources makes the technical analysis of nonlocality in networks more challenging as compared to the Bell scenario. However, networks also offer new conceptual insights.
For example, this pertains to the use of complex numbers in quantum mechanics \cite{Renou2021, Li2022,Chen2022}, device-independent certification without inputs \cite{Sekatski2022}, nonlocality with single photons~\cite{Abiuso2022}, upper bounds on measurement dependence \cite{Chaves2021} and tests of generalised probabilistic theories \cite{Weilenmann2020}.
Computational methods have been developed, mainly based on the idea of inflation, to  bound from the exterior the set of local \cite{Wolfe2019}, quantum \cite{Wolfe2021, Pozas2019} and post-quantum \cite{Gisin2020} correlations in networks.
The exploration of network nonlocality has lead to many nonlocality criteria specialised for different network architectures, for example in the bilocal scenario \cite{Branciard2010, Branciard2012, Tavakoli2021}, in the chain scenario \cite{Branciard2012, Mukherjee2015, Mukherjee2020}, in the star scenario \cite{Tavakoli2014, Tavakoli2017} and many others (see e.g.~\cite{Chaves2016, Rosset2016, Tavakoli2016, Tavakoli2016b, Luo2018, Luo2018b, Renou2022b}).

\begin{figure}
	\centering
  \begin{tikzpicture}
    \draw [fill=black] (2,3) rectangle (3,4);
    \draw (2.5, 3.5) node {\textcolor{white}{\Large $A$}};
    \draw [fill=black] (0,0) rectangle (1,1);
    \draw (0.5, 0.5) node {\textcolor{white}{\Large $B$}};
    \draw [fill=black] (4,0) rectangle (5,1);
    \draw (4.5, 0.5) node {\textcolor{white}{\Large $C$}};
    \node at (2.5, 0.5) {\Huge$*$};
    \node at (1, 2.25) {\Huge$*$};
    \node at (4, 2.25) {\Huge$*$};
    \draw [-{Latex[length=3mm]}] (2.7, 0.5) -- (4, 0.5);
    \draw [{Latex[length=3mm]}-] (1, 0.5) -- (2.3, 0.5);
    \draw [-{Latex[length=3mm]}] (0.92, 2.1) -- (0.5, 1);
    \draw [-{Latex[length=3mm]}] (1.09, 2.45) -- (2, 3.5);
    \draw [{Latex[length=3mm]}-] (3, 3.5) -- (3.88, 2.45);
    \draw [-{Latex[length=3mm]}] (4.08, 2.1) -- (4.5, 1);
    \draw [->] (0.5, 0) -- (0.5, -0.3);
    \node at (0.5, -0.55) {$b=\pm 1$};
    \draw [->] (4.5, 0) -- (4.5, -0.3);
    \node at (4.5, -0.55) {$c=\pm 1$};
    \draw [->] (2.5, 3) -- (2.5, 2.7);
    \node at (2.5, 2.45) {$a=\pm 1$};
  \end{tikzpicture}
	\caption{The simplest triangle scenario. Alice, Bob and Charlie are pairwise connected by sources that each emit a pair of particles. They each perform a selected measurement and obtain binary outcomes.}
  \label{ScenarioFig}
\end{figure}
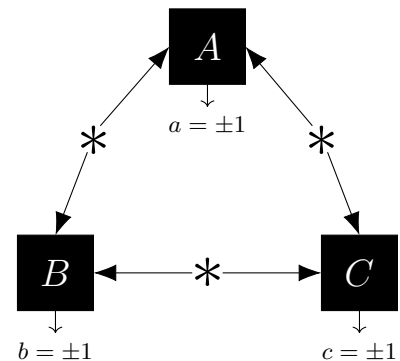

A particularly enigmatic network is the so-called triangle scenario.
It features three parties, Alice, Bob and Charlie, and three sources that each emit a pair of particles between a pair of the parties (see Figure~\ref{ScenarioFig}).
The reason that this network is of particular interest is that it is the simplest scenario in which every party is connected to every other party through a shared source.
It may be thought of as the simplest instance of a fully connected graph, in which the vertices represent the parties and the edges represent sources, each emitting independent particle pairs shared between them.
It is possible to create nonlocality in the triangle scenario by means of standard Bell nonlocality: this can be done either by directly establishing bipartite Bell nonlocal correlations between two of the parties, or by disguising them in no-input, four-outcome distributions \cite{Fritz2012}.
However, more genuine forms of triangle nonlocality are possible when none of the parties have an input and all of them give one of four possible outputs \cite{Renou2019,Pozas2023}.
This has raised the question of identifying a minimal example of nonlocality in the triangle scenario, i.e.~a probability distribution arising between Alice, Bob and Charlie that admits no local model in a scenario when they all have no input and their output alphabets are as small as possible.
It has already been found that a smaller nonlocality example is possible: there exists a quantum nonlocal probability distribution when two parties have ternary outputs and one party has a binary output \cite{Boreiri2022}.
However, the conspicuous question still remains open: is nonlocality possible in the \textit{minimal} scenario, namely when all parties have binary outputs? When restricting nonlocality to quantum theory, a negative answer has been conjectured \cite{Fraser2018}.

Here, we prove that there exists nonlocal correlations in the minimal triangle scenario that are compatible with the minimal physical requirements of no-signaling and independence (NSI) of the sources in the network, but cannot be created neither by measurements on quantum states nor by post-processing of classical randomness.
Such NSI correlations may be thought of as the network counterparts to the no-signaling correlations in Bell scenarios.
In Ref.~\cite{Gisin2020}, inflation methods were used to compute outer bounds on the set of NSI correlations in the minimal triangle scenario.
Subsequently, Ref.~\cite{Bancal2021} constructed an explicit example of a network nonlocal box that saturates a point on the NSI correlation boundary, thus proving it to be an optimal characterisation.\footnote{We note that a confusion is introduced in Ref.~\cite{Bancal2021} above equation (22), where $n\geq 2$ should read $n\geq 4$.}
Our proposal for an analogue of a Popescu-Rohrlich box (originally defined in Ref.~\cite{Popescu1994}) for networks is based on this network nonlocal box.

We here employ the inflation technique for classical correlations in networks \cite{Wolfe2019} to prove that there exists a noise-robust gap between the NSI-set and the local set of correlations in the minimal triangle scenario.
Then, we go further and investigate the same problem using quantum inflation \cite{Wolfe2021} and prove the existence of a noise-robust gap between the NSI-set and the quantum set of correlations.
Since our bounds on the quantum set and the local set do not coincide, they raise the question of whether a quantum nonlocal examplle also is possible.
However, we have been unable to find such an example.

\section{Symmetric distributions in the minimal triangle}
The distributions we consider are arguably the simplest that are supported in the triangle scenario with binary outcomes.
Namely, we consider tripartite probability distributions that are invariant under permutation of parties.
These distributions are characterised by only three independent parameters, that can be taken to be the single-, two- and three-body correlators (respectively denoted $E_1$, $E_2$ and $E_3$).
Using them, any tripartite distribution for binary outcomes that is invariant under permutation of parties can be written as
\begin{equation}
  p(a,b,c) = \frac{1}{8}[1+(a+b+c)E_1 + (ab+ac+bc)E_2 + abc E_3],
  \label{eq:prob}
\end{equation}
where $a$, $b$, $c\in\{-1,1\}$ represent the outcomes of Alice, Bob and Charlie respectively.

These distributions, and which values of $E_1$, $E_2$ and $E_3$ lead to distributions (in)compatible with NSI, were extensively studied in Ref.~\cite{Gisin2020}.
There, several analytic inequalities were derived that are satisfied by all distributions that can, in principle, be generated in the triangle scenario (this is, when the sources distribute independent classical randomness, entangled quantum states, or more exotic no-signaling systems).
In some cases, these inequalities were shown to be tight by finding concrete realisations in terms of classical shared randomness, i.e. of the form
\begin{equation}
  \begin{aligned}
    p(a,b,c) = &\sum_{\lambda_{AB}}\sum_{\lambda_{BC}}\sum_{\lambda_{AC}}p(\lambda_{AB})p(\lambda_{BC})p(\lambda_{AC}) \\
    &\times p(a|\lambda_{AB},\lambda_{AC})p(b|\lambda_{AB},\lambda_{BC})p(c|\lambda_{BC},\lambda_{AC}),
  \end{aligned}
  \label{eq:trianglelocal}
\end{equation}
which means that all distributions in those boundaries can be created by means of classical shared randomness distributed by the sources, and thus they automatically satisfy NSI.
However, in other situations, classical realisations that matched the bounds could not be found.
These are, notably, a small region in the rectangle $(E_1, E_2)\in[0.18,1/2]\times[-1/3,0]$ and a whole section in $(E_2, E_3)\in[0.3621,\sqrt{2}-1]\times[0, 0.6]$ when $E_1\,{=}\,0$.
It is in these regions where interesting phenomena may appear: there may exist distributions that, albeit being compatible with no-signaling and independence, may not be possible to realise by post-processing shared randomness, or by measuring quantum states.
We address these questions with the aid of the inflation technique \cite{Wolfe2019,Wolfe2021}.

\section{Inflation}
Inflation is a general framework for characterising correlations in causal structures (among which networks constitute a particularly interesting class). The key concept in inflation is the \textit{gedankenexperiment} of considering that one has access to multiple copies of the sources of physical systems and measurement devices that make up the network, and can arrange them in arbitrary configurations.
While at first it seems that analysing the correlations that are generated in these new configurations is an equally (if not more) daunting task, the fact that the elements in the inflation are copies of those in the original network imply a number of symmetries that
allows to simplify the characterisation, which can then be formulated in terms of linear \cite{Wolfe2019} or semidefinite programming \cite{Wolfe2021}.
In the following we provide a brief, high-level overview of the reasonings behind inflation and how different variants constrain different types of correlations, referring the reader to the original references, Refs.~\cite{Wolfe2019,Wolfe2021,Gisin2020}, for more detailed descriptions.

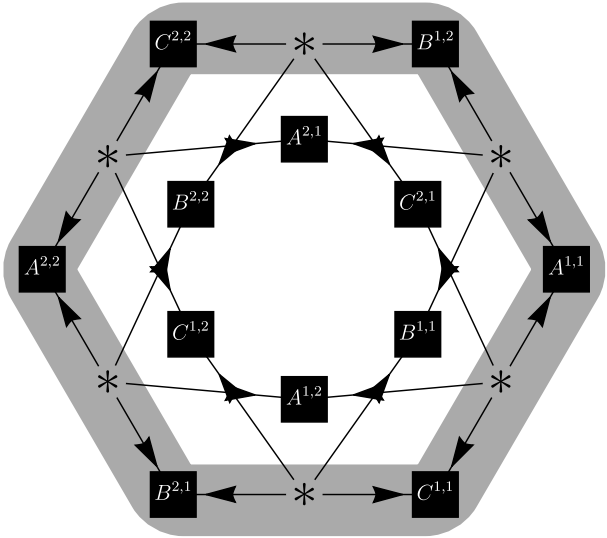
\begin{figure}
  \centering
  \begin{tikzpicture}
    \coordinate(b); 
    \foreach[count=\n from 0] \i in {0,...,5}
        \path (60*\i:2.72) coordinate (c\i) ($(c\i)!\n/(\n+1)!(b)$) coordinate(b);

    \foreach \i in {0,...,5} \node at (c\i) {};

    \draw[rounded corners=5,s/.style={scale around={1.3:(b)}},line width=9.5mm,color=plotgray]
      ([s]c0) foreach \i in{1,...,5}{--([s]c\i)}--cycle;
    \node at (0,0) {\includegraphics[width=0.9\columnwidth]{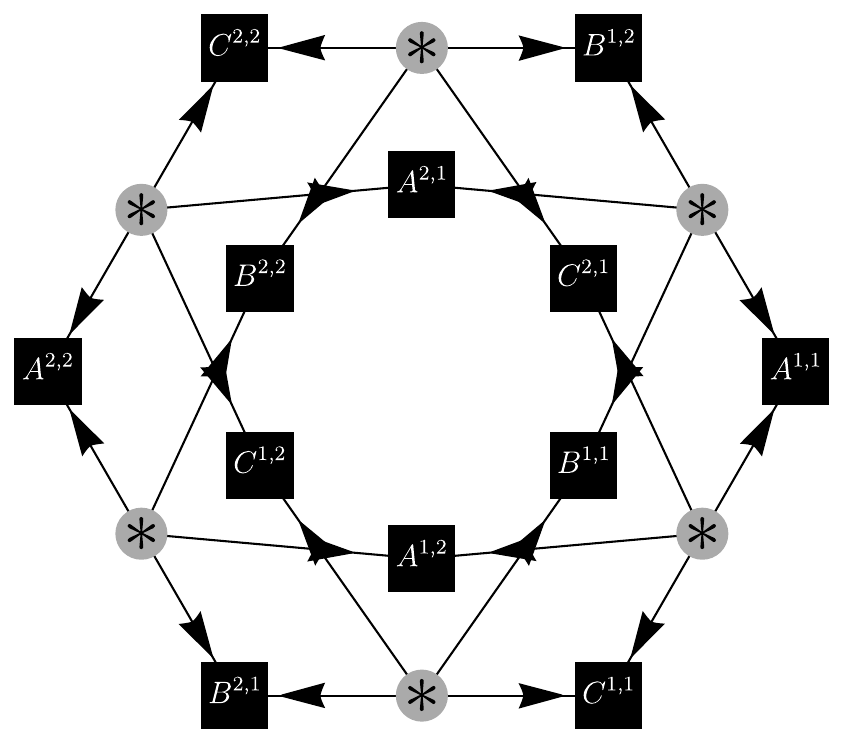}};
  \end{tikzpicture}
  \caption{The hexagon-web inflation of the triangle scenario. There are two copies of each of the sources in the original triangle scenario, and four copies of each of the parties, one for each combination of copies of sources that send systems to the corresponding party. Note that this inflation assumes that information originating from the sources can be cloned, and thus it constrains triangle-local models. The sub-network shaded is the hexagon inflation, used for instance in Ref.~\cite{Gisin2020}. Since in this case the sources do not make copies of the information sent, this inflation constrains NSI correlations.}
  \label{fig:inflation}
\end{figure}

Depending on the properties of the physical systems distributed by the sources, different types of inflations will be allowed (see Fig.~\ref{fig:inflation}, also \cite[Figs.~7, 8, 11]{ReviewPaper}).
When the sources distribute shared randomness, this can be cloned, and thus the inflations considered can include, in addition to copies of the sources and the measurement devices, copies of the shared randomness distributed to a given party.
This leads to inflations such as the hexagon-web inflation in Fig.~\ref{fig:inflation} (see also, e.g.,~\cite[Fig.~2]{Wolfe2019} or \cite[Fig.~S2]{Pozas2023}), where for a given number of copies of each source there exist as many copies of the measurement devices as possible combinations of sources distributing states.
If the sources distribute quantum systems, copying individual subsystems is prohibited due to the no-cloning theorem \cite{nocloning}, and thus the allowed inflations only contain a given number of copies of each source, and each party has one different set of measurement operators per possible combination of copies received (see, e.g.,~\cite[Fig.~4]{Wolfe2021}).
Finally, if one wants to constrain the distributions that can be generated in a given network with any type of systems (i.e., only constrained by no-signaling and independence \cite{Gisin2020}), then the inflations that must be considered are those where the information from the sources is not cloned, and each party only receives one copy of each relevant source.
This leads to inflations such as the hexagon inflation, which is the shaded sub-network in Fig.~\ref{fig:inflation}.

In addition to this, in order to make the programs as constraining as possible, we follow Ref.~\cite{Pozas2023,AlexThesis} and add, in particular, the \textit{linearised polynomial identification} (LPI) constraints that associate a factorising probability in the inflation to the product of a probability in the original network (therefore, a known, real number once $p(a,b,c)$ is fixed) and another probability that will remain as a variable in the corresponding optimisation problem.
For example, one of these constraints for distributions compatible with the inflation of Fig.~\ref{fig:inflation} is $p_\text{inf}(a^{1,1},a^{2,2},b^{2,1},c^{2,2})=p(a^{1,1})p_\text{inf}(a^{2,2},b^{2,1},c^{2,2})$.
Adding these constraints turns out to be crucial to observe the non-trivial phenomenology that we describe in the following section.

\section{Results}
\begin{figure*}
  \centering
  \subfloat[\label{fig:E1E2hexagon}]{
    \includegraphics[width=0.45\textwidth]{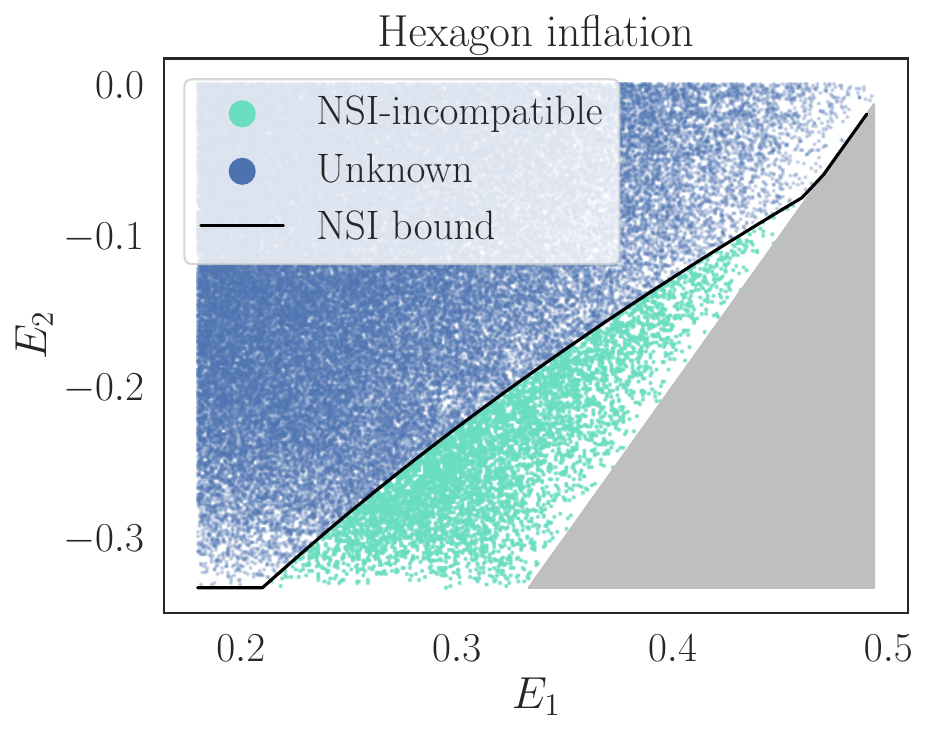}}
  \hskip 3em
  \subfloat[\label{fig:E1E2web}]{
    \includegraphics[width=0.45\textwidth]{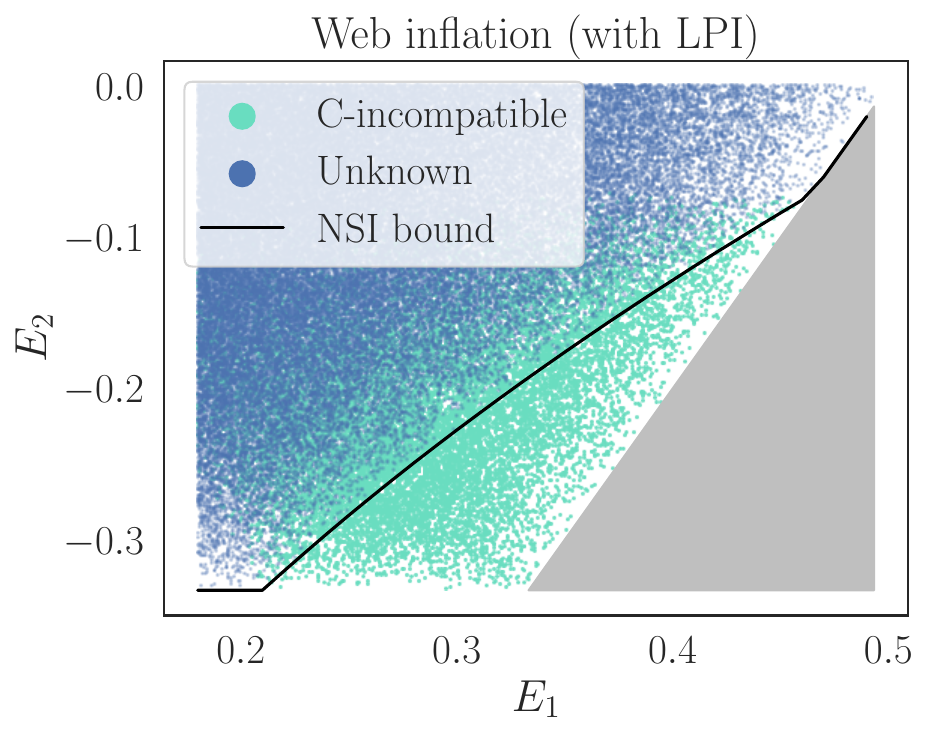}}
  \caption{Random two-outcome distributions, parameterised by $E_1$, $E_2$ and $E_3$, in the region of interest of \cite[Fig.~2]{Gisin2020}. The turquoise points denote the distributions that are incompatible with \protect\subref{fig:E1E2hexagon} the hexagon inflation, i.e., with any triangle model, and \protect\subref{fig:E1E2web} the hexagon-web inflation, i.e., with any triangle-local model. The grey region represents that for which the corresponding values of $E_1$ and $E_2$ do not produce a valid probability distribution. Notably, when the LPI constraints are not added to the classical hexagon-web inflation, the resulting figure is exactly the same as in \protect\subref{fig:E1E2hexagon}.
  \label{fig:E1E2}
  }
\end{figure*}

In the following we analyse the compatibility of the distributions in the ``interesting'' regions of the parameter space as identified by Ref.~\cite{Gisin2020} (i.e., the areas colored in yellow in \cite[Figs. 2, 3]{Gisin2020}) with classical and quantum models in the triangle network.
Classical and NSI inflation are implemented in MATLAB, and quantum inflation is implemented Python using the package \texttt{inflation} \cite{Pozas2022c}. All codes are available in the computational appendix \cite{compapp}.

We begin by focusing on the $E_1-E_2$ plane.
There, Ref.~\cite{Gisin2020} provided an inequality satisfied by any probability distribution generated in the triangle scenario, which we recover using NSI inflation in Fig.~\ref{fig:E1E2hexagon}.
Indeed, we confirm that whether a distribution can be generated in the triangle scenario can be discerned exclusively by analysing the single- and two-body marginals of the distribution and, perhaps surprisingly, ignoring the three-body correlator.
In fact, when considering which distributions do not admit a hexagon-web inflation (i.e., we ask which of those distributions can not be generated when the sources distribute classical randomness) and not imposing the LPI constraints, the perspective offered is the same as that shown in Fig.~\ref{fig:E1E2hexagon}.

\begin{figure}
  \centering
  \begin{overpic}[width=0.95\columnwidth,clip,trim={0 7cm 0 8cm}]{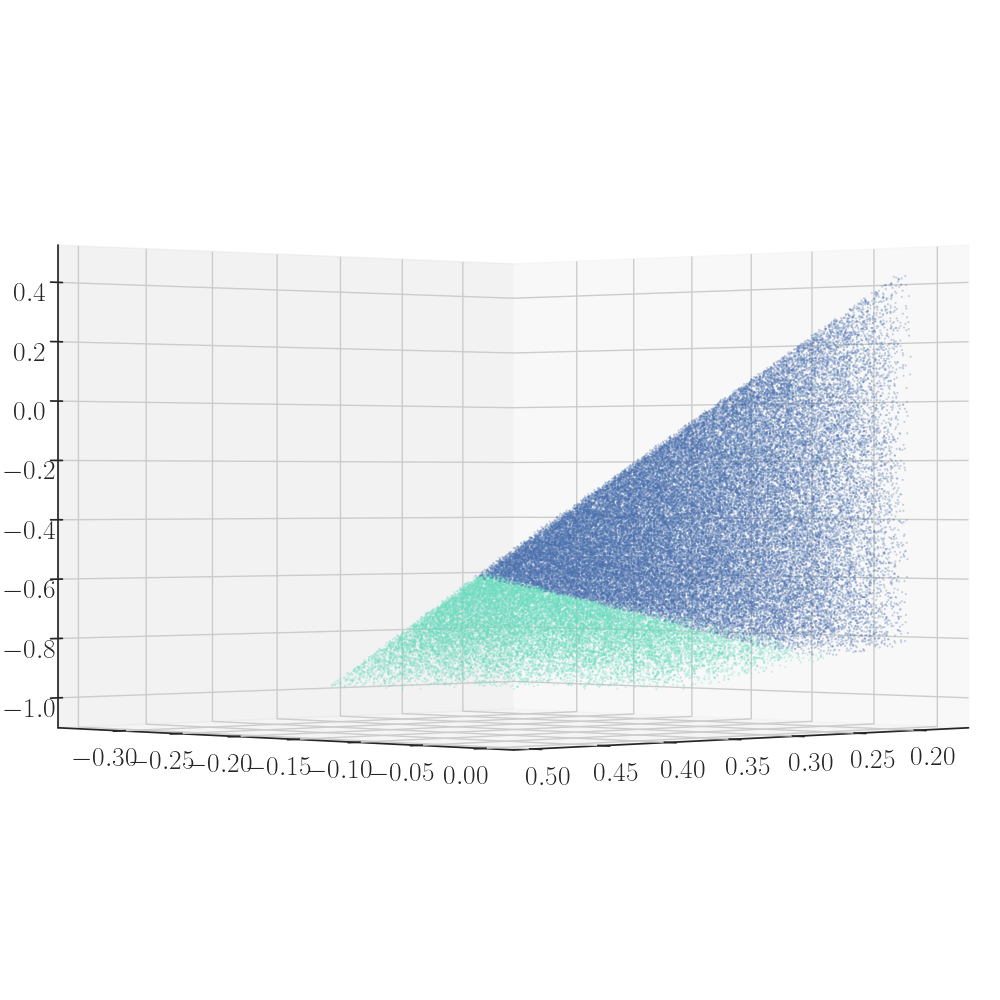}
    \put(80,0){\footnotesize $E_1$}
    \put(20,0){\footnotesize $E_2$}
    \put(-3,30){\footnotesize $E_3$}
  \end{overpic}
  \caption{Rotation of Fig.~\ref{fig:E1E2web} to depict the boundary between distributions not admitting (in light blue) and not known to admit (in dark blue) triangle-local models. The boundary can not be described by a simple polynomial in the variables $E_1$, $E_2$, $E_3$ up to degree 5.}
  \label{fig:rotated}
\end{figure}

When inserting these constraints, the picture changes drastically (see Fig.~\ref{fig:E1E2web}): we find that the three-body correlator $E_3$ has a role in determining whether the distribution $p(a,b,c)$ admits a triangle-local model of the form of Eq.~\eqref{eq:trianglelocal}.
Interestingly, we observe a non-trivial relation, since we observe that, even very close to the analytical boundary for NSI correlations of Ref.~\cite{Gisin2020}, there exist distributions that cannot be identified as not admiting a triangle-local model using the inflations considered in this work.
Changing the perspective reveals a much sharper boundary (see Fig.~\ref{fig:rotated}).
However, this boundary does not have a simple expression in terms of a polynomial in the variables $E_1$, $E_2$ and $E_3$ up to degree 5.
Numerical searches based on LHV-Net \cite{Krivachy2020} suggest that triangle-local models do not exist in the region where inflation can not guarantee the absence of a model.
It is also interesting to see that there exist distributions far from the boundary (in the area around $E_1=0.2$, $E_2=-0.1$) that are triangle nonlocal.
When visualizing the datapoints of Fig.~\ref{fig:E1E2web} in the three-dimensional space where the axes correspond to $E_1$, $E_2$ and $E_3$, a sharp boundary appears (see the computational appendix \cite{compapp} and Figure~\ref{fig:rotated}).
This boundary, despite its seemingly almost-linear form, does not have a simple expression in terms of a polynomial in the variables $E_1$, $E_2$ and $E_3$ up to degree 5.

Now we turn to the ``interesting region'' in the $E_2-E_3$ plane of \cite[Fig. 3]{Gisin2020} with the additional constraint that $E_1$ is fixed to be $0$.
In this region, Ref.~\cite{Bancal2021} showed that for each correlation with $E_1=0$ and $E_2=\sqrt{2}-1$ there exists a unique non-local box that realises it.
Positivity constraints bound the magnitude of $E_3$ by $2-\sqrt{2}$ and, as shown in \cite[Supplementary Note 2]{Gisin2020}, in the extreme values of $E_3$ the corresponding distributions admit local models. The codes implementing the classical inflation shown in Appendix~\ref{app:322inf} and the corresponding quantum inflation \cite{Wolfe2021}, both available in the computational appendix \cite{compapp}, allow to demonstrate that, up to numerical precision, $\pm(2-\sqrt{2})$ are the only values of $E_3$ for which the corresponding distribution admits a description in terms of sources of shared randomness or of quantum states.

The $E_2-E_3$ plane with $E_1$ fixed to 0 also contains an even simpler set of correlations, namely the set of noisy GHZ distributions
\begin{equation}
  p_v^{\textsc{ghz}}(a,b,c)= v \left( \frac{1}{2} \text{ if } a=b=c \right) + \frac{1-v}{8}.
  \label{eq:ghz}
\end{equation}
This set of correlations is, in addition to invariant under permutation of parties, also invariant under permutations of outcomes, and is characterised by $E_1=E_3=0$, $E_2=v$.
$E_2=v$ can be seen as the probability of the three parties sucessfully simulating a shared random bit by only using bipartite resources.
Moreover, this region is interesting because it contains what can be regarded as an equivalent of the well-known Popescu-Rohrlich box \cite{Popescu1994} for networks: namely, it is known that the distribution corresponding to $E_1=E_3=0$, $E_2=\sqrt{2}-1$, can be generated by wiring three copies of a unique non-signaling box forming a triangle \cite{Bancal2021}.
Thus, a natural question is to determine whether this distribution can equivalently admit a description in terms of sources distributing shared randomness or quantum states.

When considering the hexagon-web inflation (with LPI constraints, so the linear program used is as constraining as possible), it is possible to show that a distribution compatible with the inflation and that satisfies all the necessary constraints exists.
Therefore, the hexagon-web inflation cannot detect whether the distribution with $E_1\,{=}\,E_3\,{=}\,0$, $E_2\,{=}\,\sqrt{2}-1$ can be generated or not with sources of classical shared randomness.
However, it is not possible to find such a distribution if considering the inflation corresponding to having three copies of any of the sources in the scenario and two copies of the remaining ones, depicted in Appendix \ref{app:322inf}.
This impossibility is a rigorous proof that the distribution with $E_1=E_3=0$, $E_2=\sqrt{2}-1$ cannot be generated via sources of shared randomness.
In the computational appendix \cite{compapp} we provide a program that detects this infeasibility.

The codes in the computational appendix allow to moreover show that any distribution with $E_1=E_3=0$ is triangle nonlocal if $E_2\geq 0.3942$.
We want to note that this critical value is, a priori, not tight, and can possibly be lowered by means of considering more restricting (and thus larger and computationally more expensive) inflations.
The constructions of explicit local models in \cite{Gisin2020,Dasilva2023} strongly suggest that, in fact, the boundary between triangle-local and triangle-nonlocal distributions lies at $E_2\approx 0.3621$.
We return to this aspect later on.

In order to discern whether the distribution with $E_1\,{=}\,E_3\,{=}\,0$, $E_2\,{=}\,\sqrt{2}-1$ can indeed be considered as a network analogue of the Popescu-Rohrlich box, one would like to prove that it also does not admit a realisation in terms of sources distributing quantum systems.
In order to do so, we implement the quantum inflation \cite{Wolfe2021} corresponding to the setting above (i.e., that with three copies of one of the sources and two copies of the remaining ones, whose classical inflation is in Appendix~\ref{app:322inf}).
It is possible to show numerically that any distribution with $E_1=E_3=0$ can not be generated in the quantum triangle for $E_2\geq 0.4067$ (we provide the corresponding codes in the computational appendix \cite{compapp}).
With this, we prove that the maximally network nonlocal distribution with $E_1\,{=}\,E_3\,{=}\,0$, $E_2\,{=}\,\sqrt{2}-1$ is indeed a network analogue of the bipartite Popescu-Rohrlich box, in that it generates correlations that are stronger than those achievable in classical and quantum mechanics while still satisfying the conditions of NSI.
As in the case of classical inflation, the critical value for $E_2$ provided may not be tight.
In this case, there are two possible ways of potentially achieving stricter values \cite{Wolfe2021}: either increasing the size of the inflation as in the classical case, or increasing the level of the noncommutative polynomial optimization (NPO) hierarchy \cite{npa} associated to a given inflation.

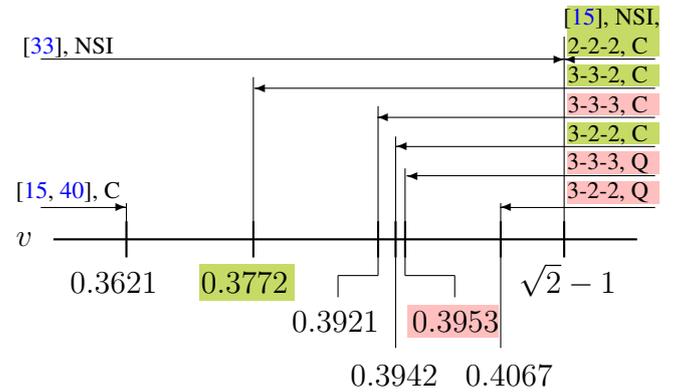
\begin{figure}
	\begin{tikzpicture}[scale=0.96]
    \node at (-4.4, 0) {\large$v$};
    \draw[thick] (-4,0) -- (4,0);
    \draw[thick] (0.69, -0.25) -- (0.69, 0.25);
    \draw[thin] (0.69, 0) -- (0.69, -1.5);
		\put(2,-55){\large$0.3942$};
    \draw[thin] (0.69, 0) -- (0.69, 1.42);
		\put(116,35){\vector(-1,0){97}};
    \fill[SpringGreen] (3.04, 1.32) rectangle (4.3,1.62);
    \node at (3.6,1.45) {3-2-2, C};

    \draw[thick] (0.45, -0.25) -- (0.45, 0.25);
    \draw[thin] (0.45, 0) -- (0.45, -0.5);
    \draw (0.45, -0.5) -- (-0.1, -0.5);
    \draw (-0.1, -0.5) -- (-0.1, -0.8);
		\put(-20,-35){\large$0.3921$};
    \draw[thin] (0.45, 0) -- (0.45, 1.84);
		\put(116,46){\vector(-1,0){104}};
    \fill[pink] (3.04, 1.72) rectangle (4.3,2.02);
    \node at (3.6,1.85) {3-3-3, C};

    \draw[thick] (-1.26, -0.25) -- (-1.26, 0.25);
    \fill[SpringGreen] (-2, -0.85) rectangle (-0.7,-0.35);
		\put(-54,-20){\large$0.3772$};
    \draw[thin] (-1.26, 0) -- (-1.26, 2.24);
		\put(116,57){\vector(-1,0){150}};
    \fill[SpringGreen] (3.04, 2.12) rectangle (4.3,2.42);
    \node at (3.6,2.25) {3-3-2, C};
    \draw[thick] (2.13, -0.25) -- (2.13, 0.25);
    \draw[thin] (2.13, 0) -- (2.13, -1.5);
		\put(45,-55){\large$0.4067$};
    \draw[thin] (2.13, 0) -- (2.13, 0.5);
		\put(116,12){\vector(-1,0){58}};
    \fill[pink] (3.04, 0.49) rectangle (4.3,0.8);
    \node at (3.6,0.65) {3-2-2, Q};

    \draw[thick] (0.82, -0.25) -- (0.82, 0.25);
    \draw[thin] (0.82, 0) -- (0.82, -0.5);
    \draw[thin] (0.82, -0.5) -- (1.5, -0.5);
    \draw[thin] (1.5, -0.5) -- (1.5, -0.8);
    \fill[pink] (0.85, -1.35) rectangle (2.1,-0.9);
		\put(25,-35){\large$0.3953$};
    \draw[thin] (0.82, 0) -- (0.82, 0.98);
		\put(116,24){\vector(-1,0){93}};
    \fill[pink] (3.04, 0.91) rectangle (4.3,1.22);
    \node at (3.6,1.05) {3-3-3, Q};
    \draw[thick] (3, -0.25) -- (3, 0.25);
		\put(65,-20){\large$\sqrt{2}-1$};
    \draw[thin] (3, 0) -- (3, 2.8);
		\put(116,68){\vector(-1,0){34}};
    \fill[SpringGreen] (3.04, 2.52) rectangle (4.3,3.22);
    \node at (3.6,2.65) {2-2-2, C};
    \node at (3.65,3.05) {\cite{Gisin2020}, NSI,};
		\put(-114,68){\vector(1,0){196}};
    \node at (-3.8,2.65) {\cite{Bancal2021}, NSI};
    \draw[thick] (-3, -0.25) -- (-3, 0.25);
		\put(-103,-20){\large$0.3621$};
    \draw[thin] (-3, 0) -- (-3, 0.5);
		\put(-114,12){\vector(1,0){32}};
    \node at (-3.8,0.65) {\cite{Dasilva2023,Gisin2020}, C};
	\end{tikzpicture}
  \vskip 2em
  \caption{Summary of results for triangle-local and triangle-quantum models for the distributions in Eq.~\eqref{eq:ghz} in the minimal triangle scenario.
    The arrows pointing rightwards represent lower bounds via the construction of explicit models, while the arrows pointing leftwards represent upper bounds obtained via inflation. For the inflations, the green color denotes the results obtained via linear programming, and the pink color denotes the results obtained via semidefinite programming. The values highlighted are the best upper bounds for classical models (green) and quantum models (pink).}
  \label{fig:ghz}
\end{figure}

Indeed, none of the critical values provided above, namely $E_2\approx 0.3942$ for triangle-local models and $E_2\approx 0.4067$ for triangle-quantum models are tight.
By trading off the size of the inflation and the level of the associated NPO hierarchy in the quantum case, we have been able to analyse compatibility with the inflation consisting of three copies of each of the sources, where each party has a total of $3^2=9$ families of operators, each corresponding to a different choice of incoming systems to measure.
Using this inflation we are able to show that any distribution with $E_1=E_3=0$ can not be generated in the quantum triangle if $E_2\geq 0.3953$.
Moreover, it is possible to use the quantum inflation technique to address compatibility with local models, by considering that all variables in the problem commute (see \cite[Sec.~VI]{Wolfe2021}).
This allows to create semidefinite programming relaxations of the inflation linear program.
By doing so, and considering the inflation consisting of three copies of each of the sources, we see that any distribution with $E_1=E_3=0$ is triangle-nonlocal if, at least, $E_2\geq 0.3921$.
This value is lower than that obtained with classical inflation and three copies of one of the sources.
However, it is possible, with moderate computational effort, to consider the classical inflation consisting of three copies of two of the sources and two copies of the remaining one.
This gives a critical value for $E_2$ that is reduced down to $E_2\geq 0.3772$, giving a gap with the lower bound of $\sim 4\%$.
Fig.~\ref{fig:ghz} summarises all these results.

One must note that the computational requirements for implementing the linear and semidefinite programs increase significantly with the amount of copies of the sources considered.
For illustration, Table~\ref{table:comput} contains an indication of the problem sizes and solving times for the inflations considered in this work.
By comparing Fig.~\ref{fig:ghz} and Table~\ref{table:comput} one can explicitly see the tradeoff associated to using semidefinite relaxations for detecting incompatibility with network-local models: using SDP relaxations allows us to go up to three copies of each of the sources in the triangle scenario with a fraction of the variables needed for the LP, at the cost of obtaining worse bounds than those obtained when considering three copies of only one of the sources (and two copies of the remaining ones).

\begin{table}
  \begin{tabular}{lccccc}
    \multicolumn{1}{c}{Inflation} & N. parties & Type & N. vars. & Size & Time (s) \\
    \hline
    2-2-2, NSI & 6 & LP & 119 & 119 $\times$ 37 & 0.1803 \\
    2-2-2, C & 12 & LP & 4201 & 4201 $\times$ 641 & 1.0732 \\
    3-2-2, C & 16 & LP & 65973 & 65973 $\times$ 3613 & 6.2624 \\
    3-3-2, C & 21 & LP & 2099453 & 2099453 $\times$ 36849 & 3446.0 \\
    3-3-3, C & 27 & SDP & 1335 & 2026 $\times$ 2026 & 5277.9 \\
    3-2-2, Q & 16 & SDP & 3798 & 1183 $\times$ 1183 & 1289.5 \\
    3-3-3, Q & 27 & SDP & 1812 & 2026 $\times$ 2026 & 5426.9
  \end{tabular}
  \caption{Problem sizes and computation times for the inflation problems considered in the work. The classical inflations are implemented in MATLAB via CVX \cite{cvx}, while the quantum inflations are implemented in Python via the \texttt{inflation} package \cite{Pozas2022c}. The resulting LPs and SDPs are solved in Mosek 9.1.9. For LPs, the number of variables is directly read from the solver and the size of the problem is ($\text{N. vars.}\,{\times}\,\text{N. constr.}$), where the number of constraints is also directly read from the solver. For SDPs, the number of variables is directly read from the solver, and the size is the dimension of the matrix $\Gamma$ such that $\Gamma\succeq 0$. Time is calculated using MATLAB's function \texttt{timeit} for the case of the LPs and Python's \texttt{timeit} package for the SDPs, and it involves both the creation and solving of the problem.}
  \label{table:comput}
\end{table}

\section{Discussion}
Can any symmetric distribution generated in the simplest triangle network be produced with only classical shared variables?
While the analysis of Ref.~\cite{Gisin2020} showed that, if the answer to the question was in the negative, these non-classical distributions would be restricted to a quite limited region of the space of parameters, here we have shown an interesting phenomenology.
By using tools based on inflation, we have assessed the compatibility of tripartite distributions only subject to no-signaling and independence in the triangle network with classical and quantum realisations.
We have found that, indeed, there exist distributions that satisfy no-signaling and independence yet cannot be created by distributing classical randomness or quantum systems in the binary-outcome triangle scenario, thereby proving that the sets of triangle-local and triangle-quantum binary-outcome distributions are strictly contained in the corresponding set of NSI distributions.
Moreover, we have showed a much richer phenomenology than that reported in Ref.~\cite{Gisin2020}, noting that the value of the three-partite correlator, $E_3$, plays a role in determining whether distributions with fixed $E_1$ and $E_2$ is triangle-nonlocal.
Lastly, we have proven that the non-local box defined by Eq.~\eqref{eq:prob} with $E_1=E_3=0$, $E_2=\sqrt{2}-1$, can be seen as the analogous of a Popescu-Rohrlich box for networks, in that it generates maximally nonlocal correlations that cannot be produced in quantum theory.

This work raises a series of questions regarding the correlations that can be achieved in the minimal triangle.
First, the phenomenology showcased in Fig.~\ref{fig:E1E2web} motivates a deeper analysis of which are the conditions, in terms of the distribution's moments, that determine (in)compatibility with triangle-local models.
Ideally, these would come in the form of analytical inequalities that would define the boundary of the set of triangle-local distributions.
The approximation of this boundary generated from outer constructions via inflation, shown in Fig.~\ref{fig:rotated}, cannot be properly fitted to polynomials up to degree 5. In the computational appendix \cite{compapp} we store the data that defines this boundary.
While inequalities can be extracted from the certificates of infeasibility of incompatible distributions, the fact that we use LPI constraints limits their range of applicability, which can be nevertheless extended via an appropriate, but in some cases intricate, analysis \cite{Pozas2023}.

Another question that remains open is whether the set of distributions admitting triangle-local models is strictly contained in the set of those admitting triangle-quantum models \cite{Fraser2018}.
Our analysis reveals distributions that admit quantum inflations but do not admit classical inflations, maintaining the possibility that the correlation sets could be different and that a classical-quantum gap exists in the minimal triangle scenario.
However, these are, at this point, only indications since inflation considers relaxations of the sets of correlations of interest.
Therefore, the possibility that using larger inflations constrains significantly the quantum set and reveals that classical and quantum correlations are equivalent in the minimal triangle scenario can not be ruled out.
Finding that the classical and quantum bounds for $v=E_2$ in Eq.~\eqref{eq:ghz} coincide would mean that simulating a shared random bit with bipartite resources is, as the guess-your-neighbors'-input game \cite{Almeida2010}, a game without quantum advantage.

Finally, there exists the possibility that the analytical boundary in Fig.~\ref{fig:E1E2}, derived in Ref.~\cite{Gisin2020}, is not tight and can be further constrained using stronger NSI inflations.
However, we expect this to be unlikely due to the fact that there exist distributions very close to it that can not be identified as not admitting even triangle-local models.

\vspace*{2em}
\begin{acknowledgments}
  This work is supported by the Spanish Ministry of Science and Innovation MCIN/AEI/10.13039/501100011033 (CEX2019-000904-S and PID2020-113523GB-I00), the Spanish Ministry of Economic Affairs and Digital Transformation (project QUANTUM ENIA, as part of the Recovery, Transformation and Resilience Plan, funded by EU program NextGenerationEU), Comunidad de Madrid (QUITEMAD-CM P2018/TCS-4342), Universidad Complutense de Madrid (FEI-EU-22-06), the CSIC Quantum Technologies Platform PTI-001, the Swiss National Science Foundation (projects 2000021\_192244/1 and P1GEP2\_199676, and the NCCR-SwissMap), the Austrian Federal Ministry of Education via the Austrian Research Promotion Agency--FFG (flagship project FO999897481, funded by EU program NextGenerationEU), the Wenner-Gren Foundation, and the Knut and Alice Wallenberg Foundation through the Wallenberg Center for Quantum Technology (WACQT).
\end{acknowledgments}

\appendix
\section{The 3-2-2 inflation of the triangle network}\label{app:322inf}
\begin{figure}[h!]
  \includegraphics[width=0.95\columnwidth]{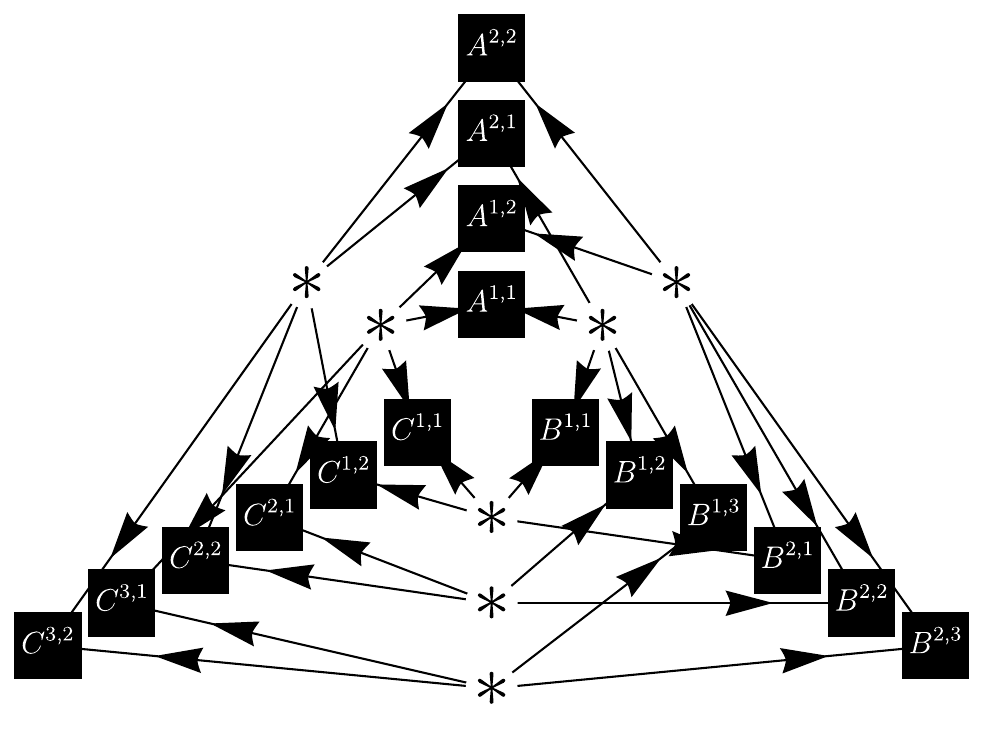}
  \caption{The 3-2-2 classical inflation of the triangle network, where there exists two copies of the sources distributing systems between parties $A$ and $B$ and between parties $A$ and $C$, and three copies of the source distributing systems between parties $B$ and $C$. Note that this inflation contains the hexagon-web inflation as a subgraph. Thus, it is a relaxation at least as powerful as the hexagon-web inflation.}
  \label{fig:322inf}
\end{figure}

Many of the results obtained in this work are obtained by considering the existence of probability distributions compatible with the inflation of the triangle network consisting of two copies of two of the sources, and three copies of the remaining one.
This inflation is depicted in Fig.~\ref{fig:322inf}.

\bibliography{references}

\begin{thebibliography}{43}%
\makeatletter
\providecommand \@ifxundefined [1]{%
 \@ifx{#1\undefined}
}%
\providecommand \@ifnum [1]{%
 \ifnum #1\expandafter \@firstoftwo
 \else \expandafter \@secondoftwo
 \fi
}%
\providecommand \@ifx [1]{%
 \ifx #1\expandafter \@firstoftwo
 \else \expandafter \@secondoftwo
 \fi
}%
\providecommand \natexlab [1]{#1}%
\providecommand \enquote  [1]{``#1''}%
\providecommand \bibnamefont  [1]{#1}%
\providecommand \bibfnamefont [1]{#1}%
\providecommand \citenamefont [1]{#1}%
\providecommand \href@noop [0]{\@secondoftwo}%
\providecommand \href [0]{\begingroup \@sanitize@url \@href}%
\providecommand \@href[1]{\@@startlink{#1}\@@href}%
\providecommand \@@href[1]{\endgroup#1\@@endlink}%
\providecommand \@sanitize@url [0]{\catcode `\\12\catcode `\$12\catcode
  `\&12\catcode `\#12\catcode `\^12\catcode `\_12\catcode `\%12\relax}%
\providecommand \@@startlink[1]{}%
\providecommand \@@endlink[0]{}%
\providecommand \url  [0]{\begingroup\@sanitize@url \@url }%
\providecommand \@url [1]{\endgroup\@href {#1}{\urlprefix }}%
\providecommand \urlprefix  [0]{URL }%
\providecommand \Eprint [0]{\href }%
\providecommand \doibase [0]{https://doi.org/}%
\providecommand \selectlanguage [0]{\@gobble}%
\providecommand \bibinfo  [0]{\@secondoftwo}%
\providecommand \bibfield  [0]{\@secondoftwo}%
\providecommand \translation [1]{[#1]}%
\providecommand \BibitemOpen [0]{}%
\providecommand \bibitemStop [0]{}%
\providecommand \bibitemNoStop [0]{.\EOS\space}%
\providecommand \EOS [0]{\spacefactor3000\relax}%
\providecommand \BibitemShut  [1]{\csname bibitem#1\endcsname}%
\let\auto@bib@innerbib\@empty
\bibitem [{\citenamefont {Tavakoli}\ \emph {et~al.}(2022)\citenamefont
  {Tavakoli}, \citenamefont {Pozas-Kerstjens}, \citenamefont {Luo},\ and\
  \citenamefont {Renou}}]{ReviewPaper}%
  \BibitemOpen
  \bibfield  {author} {\bibinfo {author} {\bibfnamefont {A.}~\bibnamefont
  {Tavakoli}}, \bibinfo {author} {\bibfnamefont {A.}~\bibnamefont
  {Pozas-Kerstjens}}, \bibinfo {author} {\bibfnamefont {M.-X.}\ \bibnamefont
  {Luo}},\ and\ \bibinfo {author} {\bibfnamefont {M.-O.}\ \bibnamefont
  {Renou}},\ }\href {https://doi.org/10.1088/1361-6633/ac41bb} {\bibfield
  {journal} {\bibinfo  {journal} {Rep. Prog. Phys.}\ }\textbf {\bibinfo
  {volume} {85}},\ \bibinfo {pages} {056001} (\bibinfo {year} {2022})},\
  \Eprint {https://arxiv.org/abs/2104.10700} {arXiv:2104.10700} \BibitemShut
  {NoStop}%
\bibitem [{\citenamefont {Branciard}\ \emph {et~al.}(2010)\citenamefont
  {Branciard}, \citenamefont {Gisin},\ and\ \citenamefont
  {Pironio}}]{Branciard2010}%
  \BibitemOpen
  \bibfield  {author} {\bibinfo {author} {\bibfnamefont {C.}~\bibnamefont
  {Branciard}}, \bibinfo {author} {\bibfnamefont {N.}~\bibnamefont {Gisin}},\
  and\ \bibinfo {author} {\bibfnamefont {S.}~\bibnamefont {Pironio}},\ }\href
  {https://doi.org/10.1103/PhysRevLett.104.170401} {\bibfield  {journal}
  {\bibinfo  {journal} {Phys. Rev. Lett.}\ }\textbf {\bibinfo {volume} {104}},\
  \bibinfo {pages} {170401} (\bibinfo {year} {2010})},\ \Eprint
  {https://arxiv.org/abs/0911.1314} {arXiv:0911.1314} \BibitemShut {NoStop}%
\bibitem [{\citenamefont {Branciard}\ \emph {et~al.}(2012)\citenamefont
  {Branciard}, \citenamefont {Rosset}, \citenamefont {Gisin},\ and\
  \citenamefont {Pironio}}]{Branciard2012}%
  \BibitemOpen
  \bibfield  {author} {\bibinfo {author} {\bibfnamefont {C.}~\bibnamefont
  {Branciard}}, \bibinfo {author} {\bibfnamefont {D.}~\bibnamefont {Rosset}},
  \bibinfo {author} {\bibfnamefont {N.}~\bibnamefont {Gisin}},\ and\ \bibinfo
  {author} {\bibfnamefont {S.}~\bibnamefont {Pironio}},\ }\href
  {https://doi.org/10.1103/PhysRevA.85.032119} {\bibfield  {journal} {\bibinfo
  {journal} {Phys. Rev. A}\ }\textbf {\bibinfo {volume} {85}},\ \bibinfo
  {pages} {032119} (\bibinfo {year} {2012})},\ \Eprint
  {https://arxiv.org/abs/1112.4502} {arXiv:1112.4502} \BibitemShut {NoStop}%
\bibitem [{\citenamefont {\ifmmode~\dot{Z}\else \.{Z}\fi{}ukowski}\ \emph
  {et~al.}(1993)\citenamefont {\ifmmode~\dot{Z}\else \.{Z}\fi{}ukowski},
  \citenamefont {Zeilinger}, \citenamefont {Horne},\ and\ \citenamefont
  {Ekert}}]{Zukowski1993}%
  \BibitemOpen
  \bibfield  {author} {\bibinfo {author} {\bibfnamefont {M.}~\bibnamefont
  {\ifmmode~\dot{Z}\else \.{Z}\fi{}ukowski}}, \bibinfo {author} {\bibfnamefont
  {A.}~\bibnamefont {Zeilinger}}, \bibinfo {author} {\bibfnamefont {M.~A.}\
  \bibnamefont {Horne}},\ and\ \bibinfo {author} {\bibfnamefont {A.~K.}\
  \bibnamefont {Ekert}},\ }\href {https://doi.org/10.1103/PhysRevLett.71.4287}
  {\bibfield  {journal} {\bibinfo  {journal} {Phys. Rev. Lett.}\ }\textbf
  {\bibinfo {volume} {71}},\ \bibinfo {pages} {4287} (\bibinfo {year}
  {1993})}\BibitemShut {NoStop}%
\bibitem [{\citenamefont {Renou}\ \emph {et~al.}(2021)\citenamefont {Renou},
  \citenamefont {Trillo}, \citenamefont {Weilenmann}, \citenamefont {Le},
  \citenamefont {Tavakoli}, \citenamefont {Gisin}, \citenamefont {Ac{\'i}n},\
  and\ \citenamefont {Navascu{\'e}s}}]{Renou2021}%
  \BibitemOpen
  \bibfield  {author} {\bibinfo {author} {\bibfnamefont {M.-O.}\ \bibnamefont
  {Renou}}, \bibinfo {author} {\bibfnamefont {D.}~\bibnamefont {Trillo}},
  \bibinfo {author} {\bibfnamefont {M.}~\bibnamefont {Weilenmann}}, \bibinfo
  {author} {\bibfnamefont {T.~P.}\ \bibnamefont {Le}}, \bibinfo {author}
  {\bibfnamefont {A.}~\bibnamefont {Tavakoli}}, \bibinfo {author}
  {\bibfnamefont {N.}~\bibnamefont {Gisin}}, \bibinfo {author} {\bibfnamefont
  {A.}~\bibnamefont {Ac{\'i}n}},\ and\ \bibinfo {author} {\bibfnamefont
  {M.}~\bibnamefont {Navascu{\'e}s}},\ }\href
  {https://doi.org/10.1038/s41586-021-04160-4} {\bibfield  {journal} {\bibinfo
  {journal} {Nature}\ }\textbf {\bibinfo {volume} {600}},\ \bibinfo {pages}
  {625} (\bibinfo {year} {2021})},\ \Eprint {https://arxiv.org/abs/2101.10873}
  {arXiv:2101.10873} \BibitemShut {NoStop}%
\bibitem [{\citenamefont {Li}\ \emph {et~al.}(2022)\citenamefont {Li},
  \citenamefont {Mao}, \citenamefont {Weilenmann}, \citenamefont {Tavakoli},
  \citenamefont {Chen}, \citenamefont {Feng}, \citenamefont {Yang},
  \citenamefont {Renou}, \citenamefont {Trillo}, \citenamefont {Le},
  \citenamefont {Gisin}, \citenamefont {Ac\'{\i}n}, \citenamefont
  {Navascu\'es}, \citenamefont {Wang},\ and\ \citenamefont {Fan}}]{Li2022}%
  \BibitemOpen
  \bibfield  {author} {\bibinfo {author} {\bibfnamefont {Z.-D.}\ \bibnamefont
  {Li}}, \bibinfo {author} {\bibfnamefont {Y.-L.}\ \bibnamefont {Mao}},
  \bibinfo {author} {\bibfnamefont {M.}~\bibnamefont {Weilenmann}}, \bibinfo
  {author} {\bibfnamefont {A.}~\bibnamefont {Tavakoli}}, \bibinfo {author}
  {\bibfnamefont {H.}~\bibnamefont {Chen}}, \bibinfo {author} {\bibfnamefont
  {L.}~\bibnamefont {Feng}}, \bibinfo {author} {\bibfnamefont {S.-J.}\
  \bibnamefont {Yang}}, \bibinfo {author} {\bibfnamefont {M.-O.}\ \bibnamefont
  {Renou}}, \bibinfo {author} {\bibfnamefont {D.}~\bibnamefont {Trillo}},
  \bibinfo {author} {\bibfnamefont {T.~P.}\ \bibnamefont {Le}}, \bibinfo
  {author} {\bibfnamefont {N.}~\bibnamefont {Gisin}}, \bibinfo {author}
  {\bibfnamefont {A.}~\bibnamefont {Ac\'{\i}n}}, \bibinfo {author}
  {\bibfnamefont {M.}~\bibnamefont {Navascu\'es}}, \bibinfo {author}
  {\bibfnamefont {Z.}~\bibnamefont {Wang}},\ and\ \bibinfo {author}
  {\bibfnamefont {J.}~\bibnamefont {Fan}},\ }\href
  {https://doi.org/10.1103/PhysRevLett.128.040402} {\bibfield  {journal}
  {\bibinfo  {journal} {Phys. Rev. Lett.}\ }\textbf {\bibinfo {volume} {128}},\
  \bibinfo {pages} {040402} (\bibinfo {year} {2022})},\ \Eprint
  {https://arxiv.org/abs/2111.15128} {arXiv:2111.15128} \BibitemShut {NoStop}%
\bibitem [{\citenamefont {Chen}\ \emph {et~al.}(2022)\citenamefont {Chen},
  \citenamefont {Wang}, \citenamefont {Liu}, \citenamefont {Wang},
  \citenamefont {Ying}, \citenamefont {Shang}, \citenamefont {Wu},
  \citenamefont {Gong}, \citenamefont {Deng}, \citenamefont {Liang},
  \citenamefont {Zhang}, \citenamefont {Peng}, \citenamefont {Zhu},
  \citenamefont {Cabello}, \citenamefont {Lu},\ and\ \citenamefont
  {Pan}}]{Chen2022}%
  \BibitemOpen
  \bibfield  {author} {\bibinfo {author} {\bibfnamefont {M.-C.}\ \bibnamefont
  {Chen}}, \bibinfo {author} {\bibfnamefont {C.}~\bibnamefont {Wang}}, \bibinfo
  {author} {\bibfnamefont {F.-M.}\ \bibnamefont {Liu}}, \bibinfo {author}
  {\bibfnamefont {J.-W.}\ \bibnamefont {Wang}}, \bibinfo {author}
  {\bibfnamefont {C.}~\bibnamefont {Ying}}, \bibinfo {author} {\bibfnamefont
  {Z.-X.}\ \bibnamefont {Shang}}, \bibinfo {author} {\bibfnamefont
  {Y.}~\bibnamefont {Wu}}, \bibinfo {author} {\bibfnamefont {M.}~\bibnamefont
  {Gong}}, \bibinfo {author} {\bibfnamefont {H.}~\bibnamefont {Deng}}, \bibinfo
  {author} {\bibfnamefont {F.-T.}\ \bibnamefont {Liang}}, \bibinfo {author}
  {\bibfnamefont {Q.}~\bibnamefont {Zhang}}, \bibinfo {author} {\bibfnamefont
  {C.-Z.}\ \bibnamefont {Peng}}, \bibinfo {author} {\bibfnamefont
  {X.}~\bibnamefont {Zhu}}, \bibinfo {author} {\bibfnamefont {A.}~\bibnamefont
  {Cabello}}, \bibinfo {author} {\bibfnamefont {C.-Y.}\ \bibnamefont {Lu}},\
  and\ \bibinfo {author} {\bibfnamefont {J.-W.}\ \bibnamefont {Pan}},\ }\href
  {https://doi.org/10.1103/PhysRevLett.128.040403} {\bibfield  {journal}
  {\bibinfo  {journal} {Phys. Rev. Lett.}\ }\textbf {\bibinfo {volume} {128}},\
  \bibinfo {pages} {040403} (\bibinfo {year} {2022})},\ \Eprint
  {https://arxiv.org/abs/2103.08123} {arXiv:2103.08123} \BibitemShut {NoStop}%
\bibitem [{\citenamefont {Sekatski}\ \emph {et~al.}(2023)\citenamefont
  {Sekatski}, \citenamefont {Boreiri},\ and\ \citenamefont
  {Brunner}}]{Sekatski2022}%
  \BibitemOpen
  \bibfield  {author} {\bibinfo {author} {\bibfnamefont {P.}~\bibnamefont
  {Sekatski}}, \bibinfo {author} {\bibfnamefont {S.}~\bibnamefont {Boreiri}},\
  and\ \bibinfo {author} {\bibfnamefont {N.}~\bibnamefont {Brunner}},\ }\href
  {https://doi.org/10.1103/PhysRevLett.131.100201} {\bibfield  {journal}
  {\bibinfo  {journal} {Phys. Rev. Lett.}\ }\textbf {\bibinfo {volume} {131}},\
  \bibinfo {pages} {100201} (\bibinfo {year} {2023})},\ \Eprint
  {https://arxiv.org/abs/2209.09921} {arXiv:2209.09921} \BibitemShut {NoStop}%
\bibitem [{\citenamefont {Abiuso}\ \emph {et~al.}(2022)\citenamefont {Abiuso},
  \citenamefont {Kriv\'achy}, \citenamefont {Boghiu}, \citenamefont {Renou},
  \citenamefont {Pozas-Kerstjens},\ and\ \citenamefont
  {Ac\'{\i}n}}]{Abiuso2022}%
  \BibitemOpen
  \bibfield  {author} {\bibinfo {author} {\bibfnamefont {P.}~\bibnamefont
  {Abiuso}}, \bibinfo {author} {\bibfnamefont {T.}~\bibnamefont {Kriv\'achy}},
  \bibinfo {author} {\bibfnamefont {E.-C.}\ \bibnamefont {Boghiu}}, \bibinfo
  {author} {\bibfnamefont {M.-O.}\ \bibnamefont {Renou}}, \bibinfo {author}
  {\bibfnamefont {A.}~\bibnamefont {Pozas-Kerstjens}},\ and\ \bibinfo {author}
  {\bibfnamefont {A.}~\bibnamefont {Ac\'{\i}n}},\ }\href
  {https://doi.org/10.1103/PhysRevResearch.4.L012041} {\bibfield  {journal}
  {\bibinfo  {journal} {Phys. Rev. Res.}\ }\textbf {\bibinfo {volume} {4}},\
  \bibinfo {pages} {L012041} (\bibinfo {year} {2022})},\ \Eprint
  {https://arxiv.org/abs/2108.01726} {arXiv:2108.01726} \BibitemShut {NoStop}%
\bibitem [{\citenamefont {Chaves}\ \emph {et~al.}(2021)\citenamefont {Chaves},
  \citenamefont {Moreno}, \citenamefont {Polino}, \citenamefont {Poderini},
  \citenamefont {Agresti}, \citenamefont {Suprano}, \citenamefont {Barros},
  \citenamefont {Carvacho}, \citenamefont {Wolfe}, \citenamefont {Canabarro},
  \citenamefont {Spekkens},\ and\ \citenamefont {Sciarrino}}]{Chaves2021}%
  \BibitemOpen
  \bibfield  {author} {\bibinfo {author} {\bibfnamefont {R.}~\bibnamefont
  {Chaves}}, \bibinfo {author} {\bibfnamefont {G.}~\bibnamefont {Moreno}},
  \bibinfo {author} {\bibfnamefont {E.}~\bibnamefont {Polino}}, \bibinfo
  {author} {\bibfnamefont {D.}~\bibnamefont {Poderini}}, \bibinfo {author}
  {\bibfnamefont {I.}~\bibnamefont {Agresti}}, \bibinfo {author} {\bibfnamefont
  {A.}~\bibnamefont {Suprano}}, \bibinfo {author} {\bibfnamefont {M.~R.}\
  \bibnamefont {Barros}}, \bibinfo {author} {\bibfnamefont {G.}~\bibnamefont
  {Carvacho}}, \bibinfo {author} {\bibfnamefont {E.}~\bibnamefont {Wolfe}},
  \bibinfo {author} {\bibfnamefont {A.}~\bibnamefont {Canabarro}}, \bibinfo
  {author} {\bibfnamefont {R.~W.}\ \bibnamefont {Spekkens}},\ and\ \bibinfo
  {author} {\bibfnamefont {F.}~\bibnamefont {Sciarrino}},\ }\href
  {https://doi.org/10.1103/PRXQuantum.2.040323} {\bibfield  {journal} {\bibinfo
   {journal} {PRX Quantum}\ }\textbf {\bibinfo {volume} {2}},\ \bibinfo {pages}
  {040323} (\bibinfo {year} {2021})},\ \Eprint
  {https://arxiv.org/abs/2105.05721} {arXiv:2105.05721} \BibitemShut {NoStop}%
\bibitem [{\citenamefont {Weilenmann}\ and\ \citenamefont
  {Colbeck}(2020)}]{Weilenmann2020}%
  \BibitemOpen
  \bibfield  {author} {\bibinfo {author} {\bibfnamefont {M.}~\bibnamefont
  {Weilenmann}}\ and\ \bibinfo {author} {\bibfnamefont {R.}~\bibnamefont
  {Colbeck}},\ }\href {https://doi.org/10.1103/PhysRevLett.125.060406}
  {\bibfield  {journal} {\bibinfo  {journal} {Phys. Rev. Lett.}\ }\textbf
  {\bibinfo {volume} {125}},\ \bibinfo {pages} {060406} (\bibinfo {year}
  {2020})},\ \Eprint {https://arxiv.org/abs/2003.00349} {arXiv:2003.00349}
  \BibitemShut {NoStop}%
\bibitem [{\citenamefont {Wolfe}\ \emph {et~al.}(2019)\citenamefont {Wolfe},
  \citenamefont {Spekkens},\ and\ \citenamefont {Fritz}}]{Wolfe2019}%
  \BibitemOpen
  \bibfield  {author} {\bibinfo {author} {\bibfnamefont {E.}~\bibnamefont
  {Wolfe}}, \bibinfo {author} {\bibfnamefont {R.~W.}\ \bibnamefont
  {Spekkens}},\ and\ \bibinfo {author} {\bibfnamefont {T.}~\bibnamefont
  {Fritz}},\ }\href {https://doi.org/10.1515/jci-2017-0020} {\bibfield
  {journal} {\bibinfo  {journal} {J. Causal Inference}\ }\textbf {\bibinfo
  {volume} {7}},\ \bibinfo {pages} {20170020} (\bibinfo {year} {2019})},\
  \Eprint {https://arxiv.org/abs/1609.00672} {arXiv:1609.00672} \BibitemShut
  {NoStop}%
\bibitem [{\citenamefont {Wolfe}\ \emph {et~al.}(2021)\citenamefont {Wolfe},
  \citenamefont {Pozas-Kerstjens}, \citenamefont {Grinberg}, \citenamefont
  {Rosset}, \citenamefont {Ac\'{\i}n},\ and\ \citenamefont
  {Navascu\'es}}]{Wolfe2021}%
  \BibitemOpen
  \bibfield  {author} {\bibinfo {author} {\bibfnamefont {E.}~\bibnamefont
  {Wolfe}}, \bibinfo {author} {\bibfnamefont {A.}~\bibnamefont
  {Pozas-Kerstjens}}, \bibinfo {author} {\bibfnamefont {M.}~\bibnamefont
  {Grinberg}}, \bibinfo {author} {\bibfnamefont {D.}~\bibnamefont {Rosset}},
  \bibinfo {author} {\bibfnamefont {A.}~\bibnamefont {Ac\'{\i}n}},\ and\
  \bibinfo {author} {\bibfnamefont {M.}~\bibnamefont {Navascu\'es}},\ }\href
  {https://doi.org/10.1103/PhysRevX.11.021043} {\bibfield  {journal} {\bibinfo
  {journal} {Phys. Rev. X}\ }\textbf {\bibinfo {volume} {11}},\ \bibinfo
  {pages} {021043} (\bibinfo {year} {2021})},\ \Eprint
  {https://arxiv.org/abs/1909.10519} {arXiv:1909.10519} \BibitemShut {NoStop}%
\bibitem [{\citenamefont {Pozas-Kerstjens}\ \emph {et~al.}(2019)\citenamefont
  {Pozas-Kerstjens}, \citenamefont {Rabelo}, \citenamefont {Rudnicki},
  \citenamefont {Chaves}, \citenamefont {Cavalcanti}, \citenamefont
  {Navascu\'es},\ and\ \citenamefont {Ac\'{\i}n}}]{Pozas2019}%
  \BibitemOpen
  \bibfield  {author} {\bibinfo {author} {\bibfnamefont {A.}~\bibnamefont
  {Pozas-Kerstjens}}, \bibinfo {author} {\bibfnamefont {R.}~\bibnamefont
  {Rabelo}}, \bibinfo {author} {\bibfnamefont {L.}~\bibnamefont {Rudnicki}},
  \bibinfo {author} {\bibfnamefont {R.}~\bibnamefont {Chaves}}, \bibinfo
  {author} {\bibfnamefont {D.}~\bibnamefont {Cavalcanti}}, \bibinfo {author}
  {\bibfnamefont {M.}~\bibnamefont {Navascu\'es}},\ and\ \bibinfo {author}
  {\bibfnamefont {A.}~\bibnamefont {Ac\'{\i}n}},\ }\href
  {https://doi.org/10.1103/PhysRevLett.123.140503} {\bibfield  {journal}
  {\bibinfo  {journal} {Phys. Rev. Lett.}\ }\textbf {\bibinfo {volume} {123}},\
  \bibinfo {pages} {140503} (\bibinfo {year} {2019})},\ \Eprint
  {https://arxiv.org/abs/1904.08943} {arXiv:1904.08943} \BibitemShut {NoStop}%
\bibitem [{\citenamefont {Gisin}\ \emph {et~al.}(2020)\citenamefont {Gisin},
  \citenamefont {Bancal}, \citenamefont {Cai}, \citenamefont {Remy},
  \citenamefont {Tavakoli}, \citenamefont {Zambrini~Cruzeiro}, \citenamefont
  {Popescu},\ and\ \citenamefont {Brunner}}]{Gisin2020}%
  \BibitemOpen
  \bibfield  {author} {\bibinfo {author} {\bibfnamefont {N.}~\bibnamefont
  {Gisin}}, \bibinfo {author} {\bibfnamefont {J.-D.}\ \bibnamefont {Bancal}},
  \bibinfo {author} {\bibfnamefont {Y.}~\bibnamefont {Cai}}, \bibinfo {author}
  {\bibfnamefont {P.}~\bibnamefont {Remy}}, \bibinfo {author} {\bibfnamefont
  {A.}~\bibnamefont {Tavakoli}}, \bibinfo {author} {\bibfnamefont
  {E.}~\bibnamefont {Zambrini~Cruzeiro}}, \bibinfo {author} {\bibfnamefont
  {S.}~\bibnamefont {Popescu}},\ and\ \bibinfo {author} {\bibfnamefont
  {N.}~\bibnamefont {Brunner}},\ }\href
  {https://doi.org/10.1038/s41467-020-16137-4} {\bibfield  {journal} {\bibinfo
  {journal} {Nat. Commun.}\ }\textbf {\bibinfo {volume} {11}},\ \bibinfo
  {pages} {2378} (\bibinfo {year} {2020})},\ \Eprint
  {https://arxiv.org/abs/1906.06495} {arXiv:1906.06495} \BibitemShut {NoStop}%
\bibitem [{\citenamefont {Tavakoli}\ \emph {et~al.}(2021)\citenamefont
  {Tavakoli}, \citenamefont {Gisin},\ and\ \citenamefont
  {Branciard}}]{Tavakoli2021}%
  \BibitemOpen
  \bibfield  {author} {\bibinfo {author} {\bibfnamefont {A.}~\bibnamefont
  {Tavakoli}}, \bibinfo {author} {\bibfnamefont {N.}~\bibnamefont {Gisin}},\
  and\ \bibinfo {author} {\bibfnamefont {C.}~\bibnamefont {Branciard}},\ }\href
  {https://doi.org/10.1103/PhysRevLett.126.220401} {\bibfield  {journal}
  {\bibinfo  {journal} {Phys. Rev. Lett.}\ }\textbf {\bibinfo {volume} {126}},\
  \bibinfo {pages} {220401} (\bibinfo {year} {2021})},\ \Eprint
  {https://arxiv.org/abs/2006.16694} {arXiv:2006.16694} \BibitemShut {NoStop}%
\bibitem [{\citenamefont {Mukherjee}\ \emph {et~al.}(2015)\citenamefont
  {Mukherjee}, \citenamefont {Paul},\ and\ \citenamefont
  {Sarkar}}]{Mukherjee2015}%
  \BibitemOpen
  \bibfield  {author} {\bibinfo {author} {\bibfnamefont {K.}~\bibnamefont
  {Mukherjee}}, \bibinfo {author} {\bibfnamefont {B.}~\bibnamefont {Paul}},\
  and\ \bibinfo {author} {\bibfnamefont {D.}~\bibnamefont {Sarkar}},\ }\href
  {https://doi.org/10.1007/s11128-015-0971-7} {\bibfield  {journal} {\bibinfo
  {journal} {Quantum Inf. Process.}\ }\textbf {\bibinfo {volume} {14}},\
  \bibinfo {pages} {2025} (\bibinfo {year} {2015})},\ \Eprint
  {https://arxiv.org/abs/1411.4188} {arXiv:1411.4188} \BibitemShut {NoStop}%
\bibitem [{\citenamefont {Mukherjee}\ \emph {et~al.}(2020)\citenamefont
  {Mukherjee}, \citenamefont {Paul},\ and\ \citenamefont
  {Roy}}]{Mukherjee2020}%
  \BibitemOpen
  \bibfield  {author} {\bibinfo {author} {\bibfnamefont {K.}~\bibnamefont
  {Mukherjee}}, \bibinfo {author} {\bibfnamefont {B.}~\bibnamefont {Paul}},\
  and\ \bibinfo {author} {\bibfnamefont {A.}~\bibnamefont {Roy}},\ }\href
  {https://doi.org/10.1103/PhysRevA.101.032328} {\bibfield  {journal} {\bibinfo
   {journal} {Phys. Rev. A}\ }\textbf {\bibinfo {volume} {101}},\ \bibinfo
  {pages} {032328} (\bibinfo {year} {2020})},\ \Eprint
  {https://arxiv.org/abs/2004.08231} {arXiv:2004.08231} \BibitemShut {NoStop}%
\bibitem [{\citenamefont {Tavakoli}\ \emph {et~al.}(2014)\citenamefont
  {Tavakoli}, \citenamefont {Skrzypczyk}, \citenamefont {Cavalcanti},\ and\
  \citenamefont {Ac\'{\i}n}}]{Tavakoli2014}%
  \BibitemOpen
  \bibfield  {author} {\bibinfo {author} {\bibfnamefont {A.}~\bibnamefont
  {Tavakoli}}, \bibinfo {author} {\bibfnamefont {P.}~\bibnamefont
  {Skrzypczyk}}, \bibinfo {author} {\bibfnamefont {D.}~\bibnamefont
  {Cavalcanti}},\ and\ \bibinfo {author} {\bibfnamefont {A.}~\bibnamefont
  {Ac\'{\i}n}},\ }\href {https://doi.org/10.1103/PhysRevA.90.062109} {\bibfield
   {journal} {\bibinfo  {journal} {Phys. Rev. A}\ }\textbf {\bibinfo {volume}
  {90}},\ \bibinfo {pages} {062109} (\bibinfo {year} {2014})},\ \Eprint
  {https://arxiv.org/abs/1409.5702} {arXiv:1409.5702} \BibitemShut {NoStop}%
\bibitem [{\citenamefont {Tavakoli}\ \emph {et~al.}(2017)\citenamefont
  {Tavakoli}, \citenamefont {Renou}, \citenamefont {Gisin},\ and\ \citenamefont
  {Brunner}}]{Tavakoli2017}%
  \BibitemOpen
  \bibfield  {author} {\bibinfo {author} {\bibfnamefont {A.}~\bibnamefont
  {Tavakoli}}, \bibinfo {author} {\bibfnamefont {M.-O.}\ \bibnamefont {Renou}},
  \bibinfo {author} {\bibfnamefont {N.}~\bibnamefont {Gisin}},\ and\ \bibinfo
  {author} {\bibfnamefont {N.}~\bibnamefont {Brunner}},\ }\href
  {https://doi.org/10.1088/1367-2630/aa7673} {\bibfield  {journal} {\bibinfo
  {journal} {New J. Phys.}\ }\textbf {\bibinfo {volume} {19}},\ \bibinfo
  {pages} {073003} (\bibinfo {year} {2017})},\ \Eprint
  {https://arxiv.org/abs/1702.03866} {arXiv:1702.03866} \BibitemShut {NoStop}%
\bibitem [{\citenamefont {Chaves}(2016)}]{Chaves2016}%
  \BibitemOpen
  \bibfield  {author} {\bibinfo {author} {\bibfnamefont {R.}~\bibnamefont
  {Chaves}},\ }\href {https://doi.org/10.1103/PhysRevLett.116.010402}
  {\bibfield  {journal} {\bibinfo  {journal} {Phys. Rev. Lett.}\ }\textbf
  {\bibinfo {volume} {116}},\ \bibinfo {pages} {010402} (\bibinfo {year}
  {2016})},\ \Eprint {https://arxiv.org/abs/1506.04325} {arXiv:1506.04325}
  \BibitemShut {NoStop}%
\bibitem [{\citenamefont {Rosset}\ \emph {et~al.}(2016)\citenamefont {Rosset},
  \citenamefont {Branciard}, \citenamefont {Barnea}, \citenamefont {P\"utz},
  \citenamefont {Brunner},\ and\ \citenamefont {Gisin}}]{Rosset2016}%
  \BibitemOpen
  \bibfield  {author} {\bibinfo {author} {\bibfnamefont {D.}~\bibnamefont
  {Rosset}}, \bibinfo {author} {\bibfnamefont {C.}~\bibnamefont {Branciard}},
  \bibinfo {author} {\bibfnamefont {T.~J.}\ \bibnamefont {Barnea}}, \bibinfo
  {author} {\bibfnamefont {G.}~\bibnamefont {P\"utz}}, \bibinfo {author}
  {\bibfnamefont {N.}~\bibnamefont {Brunner}},\ and\ \bibinfo {author}
  {\bibfnamefont {N.}~\bibnamefont {Gisin}},\ }\href
  {https://doi.org/10.1103/PhysRevLett.116.010403} {\bibfield  {journal}
  {\bibinfo  {journal} {Phys. Rev. Lett.}\ }\textbf {\bibinfo {volume} {116}},\
  \bibinfo {pages} {010403} (\bibinfo {year} {2016})},\ \Eprint
  {https://arxiv.org/abs/1506.07380} {arXiv:1506.07380} \BibitemShut {NoStop}%
\bibitem [{\citenamefont {Tavakoli}(2016{\natexlab{a}})}]{Tavakoli2016}%
  \BibitemOpen
  \bibfield  {author} {\bibinfo {author} {\bibfnamefont {A.}~\bibnamefont
  {Tavakoli}},\ }\href {https://doi.org/10.1103/PhysRevA.93.030101} {\bibfield
  {journal} {\bibinfo  {journal} {Phys. Rev. A}\ }\textbf {\bibinfo {volume}
  {93}},\ \bibinfo {pages} {030101} (\bibinfo {year} {2016}{\natexlab{a}})},\
  \Eprint {https://arxiv.org/abs/1510.05977} {arXiv:1510.05977} \BibitemShut
  {NoStop}%
\bibitem [{\citenamefont {Tavakoli}(2016{\natexlab{b}})}]{Tavakoli2016b}%
  \BibitemOpen
  \bibfield  {author} {\bibinfo {author} {\bibfnamefont {A.}~\bibnamefont
  {Tavakoli}},\ }\href {https://doi.org/10.1088/1751-8113/49/14/145304}
  {\bibfield  {journal} {\bibinfo  {journal} {J. Phys. A: Math. Theor.}\
  }\textbf {\bibinfo {volume} {49}},\ \bibinfo {pages} {145304} (\bibinfo
  {year} {2016}{\natexlab{b}})},\ \Eprint {https://arxiv.org/abs/1509.08491}
  {arXiv:1509.08491} \BibitemShut {NoStop}%
\bibitem [{\citenamefont {Luo}(2018{\natexlab{a}})}]{Luo2018}%
  \BibitemOpen
  \bibfield  {author} {\bibinfo {author} {\bibfnamefont {M.-X.}\ \bibnamefont
  {Luo}},\ }\href {https://doi.org/10.1103/PhysRevLett.120.140402} {\bibfield
  {journal} {\bibinfo  {journal} {Phys. Rev. Lett.}\ }\textbf {\bibinfo
  {volume} {120}},\ \bibinfo {pages} {140402} (\bibinfo {year}
  {2018}{\natexlab{a}})},\ \Eprint {https://arxiv.org/abs/1707.09517}
  {arXiv:1707.09517} \BibitemShut {NoStop}%
\bibitem [{\citenamefont {Luo}(2018{\natexlab{b}})}]{Luo2018b}%
  \BibitemOpen
  \bibfield  {author} {\bibinfo {author} {\bibfnamefont {M.-X.}\ \bibnamefont
  {Luo}},\ }\href {https://doi.org/10.1103/PhysRevA.98.042317} {\bibfield
  {journal} {\bibinfo  {journal} {Phys. Rev. A}\ }\textbf {\bibinfo {volume}
  {98}},\ \bibinfo {pages} {042317} (\bibinfo {year} {2018}{\natexlab{b}})},\
  \Eprint {https://arxiv.org/abs/1806.09758} {arXiv:1806.09758} \BibitemShut
  {NoStop}%
\bibitem [{\citenamefont {Renou}\ and\ \citenamefont
  {Beigi}(2022)}]{Renou2022b}%
  \BibitemOpen
  \bibfield  {author} {\bibinfo {author} {\bibfnamefont {M.-O.}\ \bibnamefont
  {Renou}}\ and\ \bibinfo {author} {\bibfnamefont {S.}~\bibnamefont {Beigi}},\
  }\href {https://doi.org/10.1103/PhysRevLett.128.060401} {\bibfield  {journal}
  {\bibinfo  {journal} {Phys. Rev. Lett.}\ }\textbf {\bibinfo {volume} {128}},\
  \bibinfo {pages} {060401} (\bibinfo {year} {2022})},\ \Eprint
  {https://arxiv.org/abs/2011.02769} {arXiv:2011.02769} \BibitemShut {NoStop}%
\bibitem [{\citenamefont {Fritz}(2012)}]{Fritz2012}%
  \BibitemOpen
  \bibfield  {author} {\bibinfo {author} {\bibfnamefont {T.}~\bibnamefont
  {Fritz}},\ }\href {https://doi.org/10.1088/1367-2630/14/10/103001} {\bibfield
   {journal} {\bibinfo  {journal} {New J. Phys.}\ }\textbf {\bibinfo {volume}
  {14}},\ \bibinfo {pages} {103001} (\bibinfo {year} {2012})},\ \Eprint
  {https://arxiv.org/abs/1206.5115} {arXiv:1206.5115} \BibitemShut {NoStop}%
\bibitem [{\citenamefont {Renou}\ \emph {et~al.}(2019)\citenamefont {Renou},
  \citenamefont {B\"aumer}, \citenamefont {Boreiri}, \citenamefont {Brunner},
  \citenamefont {Gisin},\ and\ \citenamefont {Beigi}}]{Renou2019}%
  \BibitemOpen
  \bibfield  {author} {\bibinfo {author} {\bibfnamefont {M.-O.}\ \bibnamefont
  {Renou}}, \bibinfo {author} {\bibfnamefont {E.}~\bibnamefont {B\"aumer}},
  \bibinfo {author} {\bibfnamefont {S.}~\bibnamefont {Boreiri}}, \bibinfo
  {author} {\bibfnamefont {N.}~\bibnamefont {Brunner}}, \bibinfo {author}
  {\bibfnamefont {N.}~\bibnamefont {Gisin}},\ and\ \bibinfo {author}
  {\bibfnamefont {S.}~\bibnamefont {Beigi}},\ }\href
  {https://doi.org/10.1103/PhysRevLett.123.140401} {\bibfield  {journal}
  {\bibinfo  {journal} {Phys. Rev. Lett.}\ }\textbf {\bibinfo {volume} {123}},\
  \bibinfo {pages} {140401} (\bibinfo {year} {2019})},\ \Eprint
  {https://arxiv.org/abs/1905.04902} {arXiv:1905.04902} \BibitemShut {NoStop}%
\bibitem [{\citenamefont {Pozas-Kerstjens}\ \emph {et~al.}(2023)\citenamefont
  {Pozas-Kerstjens}, \citenamefont {Gisin},\ and\ \citenamefont
  {Renou}}]{Pozas2023}%
  \BibitemOpen
  \bibfield  {author} {\bibinfo {author} {\bibfnamefont {A.}~\bibnamefont
  {Pozas-Kerstjens}}, \bibinfo {author} {\bibfnamefont {N.}~\bibnamefont
  {Gisin}},\ and\ \bibinfo {author} {\bibfnamefont {M.-O.}\ \bibnamefont
  {Renou}},\ }\href {https://doi.org/10.1103/PhysRevLett.130.090201} {\bibfield
   {journal} {\bibinfo  {journal} {Phys. Rev. Lett.}\ }\textbf {\bibinfo
  {volume} {130}},\ \bibinfo {pages} {090201} (\bibinfo {year} {2023})},\
  \Eprint {https://arxiv.org/abs/2203.16543} {arXiv:2203.16543} \BibitemShut
  {NoStop}%
\bibitem [{\citenamefont {Boreiri}\ \emph {et~al.}(2023)\citenamefont
  {Boreiri}, \citenamefont {Girardin}, \citenamefont {Ulu}, \citenamefont
  {Lypka-Bartosik}, \citenamefont {Brunner},\ and\ \citenamefont
  {Sekatski}}]{Boreiri2022}%
  \BibitemOpen
  \bibfield  {author} {\bibinfo {author} {\bibfnamefont {S.}~\bibnamefont
  {Boreiri}}, \bibinfo {author} {\bibfnamefont {A.}~\bibnamefont {Girardin}},
  \bibinfo {author} {\bibfnamefont {B.}~\bibnamefont {Ulu}}, \bibinfo {author}
  {\bibfnamefont {P.}~\bibnamefont {Lypka-Bartosik}}, \bibinfo {author}
  {\bibfnamefont {N.}~\bibnamefont {Brunner}},\ and\ \bibinfo {author}
  {\bibfnamefont {P.}~\bibnamefont {Sekatski}},\ }\href
  {https://doi.org/10.1103/PhysRevA.107.062413} {\bibfield  {journal} {\bibinfo
   {journal} {Phys. Rev. A}\ }\textbf {\bibinfo {volume} {107}},\ \bibinfo
  {pages} {062413} (\bibinfo {year} {2023})},\ \Eprint
  {https://arxiv.org/abs/2207.08532} {arXiv:2207.08532} \BibitemShut {NoStop}%
\bibitem [{\citenamefont {Fraser}\ and\ \citenamefont
  {Wolfe}(2018)}]{Fraser2018}%
  \BibitemOpen
  \bibfield  {author} {\bibinfo {author} {\bibfnamefont {T.~C.}\ \bibnamefont
  {Fraser}}\ and\ \bibinfo {author} {\bibfnamefont {E.}~\bibnamefont {Wolfe}},\
  }\href {https://doi.org/10.1103/PhysRevA.98.022113} {\bibfield  {journal}
  {\bibinfo  {journal} {Phys. Rev. A}\ }\textbf {\bibinfo {volume} {98}},\
  \bibinfo {pages} {022113} (\bibinfo {year} {2018})},\ \Eprint
  {https://arxiv.org/abs/1709.06242} {arXiv:1709.06242} \BibitemShut {NoStop}%
\bibitem [{\citenamefont {Bancal}\ and\ \citenamefont
  {Gisin}(2021)}]{Bancal2021}%
  \BibitemOpen
  \bibfield  {author} {\bibinfo {author} {\bibfnamefont {J.-D.}\ \bibnamefont
  {Bancal}}\ and\ \bibinfo {author} {\bibfnamefont {N.}~\bibnamefont {Gisin}},\
  }\href {https://doi.org/10.1103/PhysRevA.104.052212} {\bibfield  {journal}
  {\bibinfo  {journal} {Phys. Rev. A}\ }\textbf {\bibinfo {volume} {104}},\
  \bibinfo {pages} {052212} (\bibinfo {year} {2021})},\ \Eprint
  {https://arxiv.org/abs/2102.03597} {arXiv:2102.03597} \BibitemShut {NoStop}%
\bibitem [{\citenamefont {Popescu}\ and\ \citenamefont
  {Rohrlich}(1994)}]{Popescu1994}%
  \BibitemOpen
  \bibfield  {author} {\bibinfo {author} {\bibfnamefont {S.}~\bibnamefont
  {Popescu}}\ and\ \bibinfo {author} {\bibfnamefont {D.}~\bibnamefont
  {Rohrlich}},\ }\href {https://doi.org/10.1007/BF02058098} {\bibfield
  {journal} {\bibinfo  {journal} {Found. Phys.}\ }\textbf {\bibinfo {volume}
  {24}},\ \bibinfo {pages} {379} (\bibinfo {year} {1994})},\ \Eprint
  {https://arxiv.org/abs/quant-ph/9508009} {arXiv:quant-ph/9508009}
  \BibitemShut {NoStop}%
\bibitem [{\citenamefont {Wootters}\ and\ \citenamefont
  {Zurek}(1982)}]{nocloning}%
  \BibitemOpen
  \bibfield  {author} {\bibinfo {author} {\bibfnamefont {W.~K.}\ \bibnamefont
  {Wootters}}\ and\ \bibinfo {author} {\bibfnamefont {W.~H.}\ \bibnamefont
  {Zurek}},\ }\href {https://doi.org/10.1038/299802a0} {\bibfield  {journal}
  {\bibinfo  {journal} {Nature}\ }\textbf {\bibinfo {volume} {299}},\ \bibinfo
  {pages} {802} (\bibinfo {year} {1982})}\BibitemShut {NoStop}%
\bibitem [{\citenamefont {Pozas-Kerstjens}(2019)}]{AlexThesis}%
  \BibitemOpen
  \bibfield  {author} {\bibinfo {author} {\bibfnamefont {A.}~\bibnamefont
  {Pozas-Kerstjens}},\ }\emph {\bibinfo {title} {Quantum information outside
  quantum information}},\ \href {http://hdl.handle.net/10803/667696} {Ph.D.
  thesis},\ \bibinfo  {school} {Universitat Polit\`ecnica de Catalunya}
  (\bibinfo {year} {2019})\BibitemShut {NoStop}%
\bibitem [{\citenamefont {Boghiu}\ \emph {et~al.}(2023)\citenamefont {Boghiu},
  \citenamefont {Wolfe},\ and\ \citenamefont {Pozas-Kerstjens}}]{Pozas2022c}%
  \BibitemOpen
  \bibfield  {author} {\bibinfo {author} {\bibfnamefont {E.-C.}\ \bibnamefont
  {Boghiu}}, \bibinfo {author} {\bibfnamefont {E.}~\bibnamefont {Wolfe}},\ and\
  \bibinfo {author} {\bibfnamefont {A.}~\bibnamefont {Pozas-Kerstjens}},\
  }\href {https://doi.org/10.22331/q-2023-05-04-996} {\bibfield  {journal}
  {\bibinfo  {journal} {{Quantum}}\ }\textbf {\bibinfo {volume} {7}},\ \bibinfo
  {pages} {996} (\bibinfo {year} {2023})},\ \Eprint
  {https://arxiv.org/abs/2211.04483} {arXiv:2211.04483} \BibitemShut {NoStop}%
\bibitem [{\citenamefont {Pozas-Kerstjens}(2023)}]{compapp}%
  \BibitemOpen
  \bibfield  {author} {\bibinfo {author} {\bibfnamefont {A.}~\bibnamefont
  {Pozas-Kerstjens}},\ }\href
  {https://github.com/apozas/minimal-triangle-nonlocality} {\bibinfo {title}
  {{Computational appendix of \textit{Post-quantum nonlocality in the minimal
  triangle scenario}}}},\ \bibinfo {howpublished} {GitHub repository} (\bibinfo
  {year} {2023})\BibitemShut {NoStop}%
\bibitem [{\citenamefont {Kriv\'achy}\ \emph {et~al.}(2020)\citenamefont
  {Kriv\'achy}, \citenamefont {Cai}, \citenamefont {Cavalcanti}, \citenamefont
  {Tavakoli}, \citenamefont {Gisin},\ and\ \citenamefont
  {Brunner}}]{Krivachy2020}%
  \BibitemOpen
  \bibfield  {author} {\bibinfo {author} {\bibfnamefont {T.}~\bibnamefont
  {Kriv\'achy}}, \bibinfo {author} {\bibfnamefont {Y.}~\bibnamefont {Cai}},
  \bibinfo {author} {\bibfnamefont {D.}~\bibnamefont {Cavalcanti}}, \bibinfo
  {author} {\bibfnamefont {A.}~\bibnamefont {Tavakoli}}, \bibinfo {author}
  {\bibfnamefont {N.}~\bibnamefont {Gisin}},\ and\ \bibinfo {author}
  {\bibfnamefont {N.}~\bibnamefont {Brunner}},\ }\href
  {https://doi.org/10.1038/s41534-020-00305-x} {\bibfield  {journal} {\bibinfo
  {journal} {npj Quantum Inf.}\ }\textbf {\bibinfo {volume} {6}},\ \bibinfo
  {pages} {70} (\bibinfo {year} {2020})},\ \Eprint
  {https://arxiv.org/abs/1907.10552} {arXiv:1907.10552} \BibitemShut {NoStop}%
\bibitem [{\citenamefont {da~Silva}\ and\ \citenamefont
  {Parisio}(2023)}]{Dasilva2023}%
  \BibitemOpen
  \bibfield  {author} {\bibinfo {author} {\bibfnamefont {J.~M.}\ \bibnamefont
  {da~Silva}}\ and\ \bibinfo {author} {\bibfnamefont {F.}~\bibnamefont
  {Parisio}},\ }\href {https://doi.org/10.1103/PhysRevA.108.052602} {\bibfield
  {journal} {\bibinfo  {journal} {Phys. Rev. A}\ }\textbf {\bibinfo {volume}
  {108}},\ \bibinfo {pages} {052602} (\bibinfo {year} {2023})},\ \Eprint
  {https://arxiv.org/abs/2303.09954} {arXiv:2303.09954} \BibitemShut {NoStop}%
\bibitem [{\citenamefont {Navascués}\ \emph {et~al.}(2008)\citenamefont
  {Navascués}, \citenamefont {Pironio},\ and\ \citenamefont {Acín}}]{npa}%
  \BibitemOpen
  \bibfield  {author} {\bibinfo {author} {\bibfnamefont {M.}~\bibnamefont
  {Navascués}}, \bibinfo {author} {\bibfnamefont {S.}~\bibnamefont
  {Pironio}},\ and\ \bibinfo {author} {\bibfnamefont {A.}~\bibnamefont
  {Acín}},\ }\href {https://doi.org/10.1088/1367-2630/10/7/073013} {\bibfield
  {journal} {\bibinfo  {journal} {New J. Phys.}\ }\textbf {\bibinfo {volume}
  {10}},\ \bibinfo {pages} {073013} (\bibinfo {year} {2008})},\ \Eprint
  {https://arxiv.org/abs/0803.4290} {arXiv:0803.4290} \BibitemShut {NoStop}%
\bibitem [{\citenamefont {Grant}\ and\ \citenamefont {Boyd}(2014)}]{cvx}%
  \BibitemOpen
  \bibfield  {author} {\bibinfo {author} {\bibfnamefont {M.}~\bibnamefont
  {Grant}}\ and\ \bibinfo {author} {\bibfnamefont {S.}~\bibnamefont {Boyd}},\
  }\href@noop {} {\bibinfo {title} {{CVX}: {MATLAB} software for disciplined
  convex programming, version 2.1}},\ \bibinfo {howpublished}
  {\url{http://cvxr.com/cvx}} (\bibinfo {year} {2014})\BibitemShut {NoStop}%
\bibitem [{\citenamefont {Almeida}\ \emph {et~al.}(2010)\citenamefont
  {Almeida}, \citenamefont {Bancal}, \citenamefont {Brunner}, \citenamefont
  {Ac\'{\i}n}, \citenamefont {Gisin},\ and\ \citenamefont
  {Pironio}}]{Almeida2010}%
  \BibitemOpen
  \bibfield  {author} {\bibinfo {author} {\bibfnamefont {M.~L.}\ \bibnamefont
  {Almeida}}, \bibinfo {author} {\bibfnamefont {J.-D.}\ \bibnamefont {Bancal}},
  \bibinfo {author} {\bibfnamefont {N.}~\bibnamefont {Brunner}}, \bibinfo
  {author} {\bibfnamefont {A.}~\bibnamefont {Ac\'{\i}n}}, \bibinfo {author}
  {\bibfnamefont {N.}~\bibnamefont {Gisin}},\ and\ \bibinfo {author}
  {\bibfnamefont {S.}~\bibnamefont {Pironio}},\ }\href
  {https://doi.org/10.1103/PhysRevLett.104.230404} {\bibfield  {journal}
  {\bibinfo  {journal} {Phys. Rev. Lett.}\ }\textbf {\bibinfo {volume} {104}},\
  \bibinfo {pages} {230404} (\bibinfo {year} {2010})},\ \Eprint
  {https://arxiv.org/abs/1003.3844} {arXiv:1003.3844} \BibitemShut {NoStop}%
\end{thebibliography}%

\end{document}